# Numerical simulation of thermal noise in Josephson circuits

K Segall[1,5], D Schult[2], U Ray[1,3], and T Ohsumi[4]

[1] Department of Physics and Astronomy, Colgate University, 13 Oak Drive, Hamilton, NY 13346
[2] Department of Mathematics, Colgate University, 13 Oak Drive, Hamilton, NY 13346
[3] Department of Physics, University of Illinois at Urbana-Champaign, 1110 West Green Street, Urbana, IL 61801
[4] Department of Molecular Biology, Massachusetts General Hospital, Boston, MA 02114
[5] Author correspondence: ksegall@colgate.edu

**Abstract**
We present a method to numerically add thermal noise to the equations of motion for a circuit of Josephson junctions. A new noise term, which we call "linearly interpolated Gaussian noise," replaces the usual white noise process. It consists of random noise values spaced at a chosen time interval and linearly interpolated in-between. This method can be used with variable time step solvers, allowing more precise control over the error while ensuring that fast dynamics are not missed by the solver. We derive the spectral density of such a noise term and compare it to a white noise process with agreement below a cutoff frequency determined by the choice of time interval. Then we demonstrate the technique by computing the switching dynamics of Josephson circuits and comparing the results both to the traditional computational method and to experimental data.

# 1 Introduction

Over the past few decades there has been an ever-increasing interest in the dynamics of Josephson junctions at low temperatures. Single Josephson junctions have shown thermal activation [1], macroscopic quantum tunneling (MQT) [2] and phase diffusion [3]. Isolated circuits of 1-3 junctions acting as 2-state quantum systems have demonstrated coherent quantum dynamics and are serious candidates for the "qubits" of a future solid state quantum computer [4]. 1-D and 2-D Arrays of Josephson junctions have been used to study nonlinear phenomena such as vortices [5] and discrete breathers [6]. And finally, the detection of classical or quantum states of charge, phase or flux in Josephson (or other) systems is often done with separate Josephson junctions circuits acting as sensors. Superconducting Quantum Interference Devices (SQUIDs) are of course the standard for detecting flux [7], but also the switching current of a Josephson junction and the resonant frequency of the phase particle have been used in different detection schemes as well [8].

In most of the examples above, thermal noise plays an important role in the dynamics of the Josephson phase, the major degree of freedom for a Josephson junction. In studies of Josephson sensors, for example, thermal noise can play a key role in determining the signal-to-noise resolution of the detector. In the original work on MQT, thermal noise needed to be well-understood in order to infer the presence of quantum tunneling [2]. In more recent studies of quantum computing, thermal noise can act as a major source of decoherence for Josephson qubits [9]. In studies of nonlinear phenomena, thermal noise can act as a random "depinning" force or cause the creation of vortex-antivortex pairs [5]. Thermal noise must be taken into account at some level for most studies involving Josephson junctions.

The classical dynamics of the Josephson phase for a single Josephson junction in the RCSJ model is described by a nonlinear, pendulum-like equation [10]. This equation can only be solved exactly under special conditions. As the number of junctions in the circuit increases, numerical simulation becomes more important since the equations cannot be solved analytically. Since many of the parameters (such as the critical current, capacitance or inductance) in Josephson circuits can be independently determined, numerical simulation can be extremely accurate in comparing to experiment.

The numerical techniques used most often are the conventional Runga-Kutta methods [11] or Cash-Carp methods [12]. These take fixed steps in time, and thermal white noise can be added as a random variable at each time step. With fixed time steps the amplitude of the noise term is straightforward, since the reciprocal of the time step is proportional to the bandwidth of the noise. While these methods have certainly proved successful, they have two drawbacks. First of all, the choice of the time step represents a trade off, as shorter time steps are more accurate but lengthen the computational time. Quite often computations must be run over and over with successively shorter time steps to make sure no dynamics have been "missed." Moreover, if the critical timestep for resolution of the solution varies in time, the smallest of all these must be used.

The use of variable time step methods [11] can alleviate some of these problems. Variable time step methods change the size of the time step to guarantee convergence to within a certain error tolerance. Convergence to an absolute error level can also help ensure that stiff dynamics are not missed. However, because the time step is variable, the needed bandwidth of the noise is unknown at each time point.

In this work we present a numerical method to simulate thermal noise in a system of coupled Josephson junctions for use with a variable time step solver. A noise function is created by random sampling at a uniform pre-chosen time step $\eta$ and then linearly interpolated in-between these steps. This function can simply be added to the equation of motion for the Josephson phase. Section 2 explains the method. In section 3 we calculate the average power spectral density of such a noise function and show how it can be made equivalent to Gaussian white noise up to a chosen cut-off frequency. In section 4 we demonstrate the use of the method by calculating the switching dynamics of a 2-junction underdamped DC SQUID for different parameters. In section 5 we compare to an experimental study. We end with a discussion of the results and how to choose the timestep parameter $\eta$.

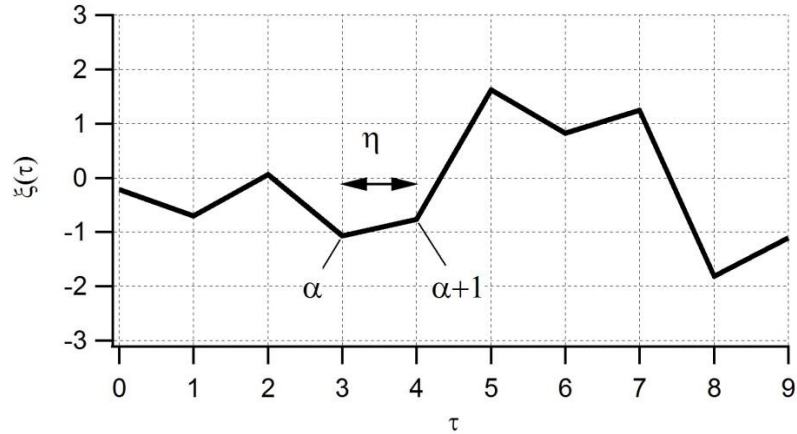

**Figure 1:** Schematic of linearly interpolated Gaussian Noise. Random values from a normal distribution are chosen at each point in time and linearly interpolated in-between. The time between samplings is uniform and here shown as *η=1*.

## 2 Method
The equation of motion of a Josephson junction at a temperature $T$ is given by [10]:

$$\ddot{\phi} = -\gamma\dot{\phi} - \sin\phi + i_J + (2\gamma\tilde{T})^{1/2}\Gamma(\tau) \tag{1}$$

Here $\phi$ is the guage-invariant phase difference across the junction, and $\gamma$ is the normalized damping parameter given by $\gamma^2 = \Phi_0/(2\pi I_C R^2 C)$, where $I_c$, $R$ and $C$ are the critical current, resistance and capacitance of the Josephson junction, respectively, and $\Phi_0$ is the flux quantum. The total current flowing through the junction, either from external supplies or from coupling to other junctions, is given by $i_J$, which has been normalized to the critical current $I_c$. The junction's normalized temperature is $\tilde{T} = 2\pi k_B T /(\Phi_0 I_c)$, where $k_B$ is Boltzmann's constant and T is the unnormalized temperature. The dot notation in equation (1) refers to differentiation with respect to the normalized time $\tau$ given by $\tau^2 = t^2 \Phi_0 C /(2\pi I_C)$, where $t$ is the unnormalized time. Finally, $\Gamma(\tau)$ refers to the Gaussian, infinite-intensity white noise process [13].

A simple scheme to solve equation (1) numerically is to define $v = \dot{\phi}$ and use the update form of equation (1) for $\dot{v} = \ddot{\phi}$ [13]:

$$v(\tau + d\tau) = v(\tau) + [\gamma v - \sin\phi + i_J]d\tau + (2\gamma\tilde{T})^{1/2}(d\tau)^{1/2}N(0,1) \tag{2}$$

Numerical simulation of thermal noise in Josephson circuitsHere $N(0,1)$ represents a sampling of the Gaussian random variable with mean equal to zero and standard deviation equal to one, and $d\tau$ is the timestep. For a fixed timestep solver, a random value is chosen for $N(0,1)$ for each timestep.

With a variable timestep solver, the implementation of the stochastic noise is not as straightforward. The adapting process sometimes requires restarting the solver from a previous point in time with a new step size. When the error estimate is too large, the noise bandwidth scaling parameter must be adjusted with the timestep in order to be consistent with the underlying stochastic process. This has the potential to cause a time correlation in the noise terms when the timesteps are adjusted.

We alleviate this problem by discretizing the stochastic process separately from the equation of motion. We use a uniform step size $\eta$ for the stochastic process, interpolating linearly between discrete points if needed. In effect, we replace the white noise process $\Gamma(t)$ in equation (1) with a new term which we call *linearly interpolated Gaussian noise*. It is denoted $\xi(\tau)$ and is given by:

$$\xi(\tau) = \frac{[N_{\alpha+1}(0,1) - N_\alpha(0,1)]}{\eta}(\tau - \alpha\eta) + N_\alpha(0,1) \tag{3}$$

where $\alpha = \text{floor}(\tau/\eta)$ is an integer and the subscript $\alpha$ on $N_\alpha(0,1)$ represents the $\alpha^{th}$ sampling of the random normal distribution. As defined, $\xi(\tau)$ has variance 1. When implemented, a scaling factor controls the variance so that $\sigma\xi(\tau)$ has variance $\sigma^2$. To mimic white noise we should pick $\sigma = (1/\eta)^{1/2}$. Figure 1 shows a schematic of one realization of $\xi(\tau)$. It is a set of random values spaced by an amount $\eta$ in time, with linear interpolation in-between to form a continuous function. Equation (3) can alternatively be written as

$$\xi(\tau) = m_\alpha(\tau - \alpha\eta) + c_\alpha \tag{4}$$

where $m_\alpha$ and $c_\alpha$ determine the slope and intercept of the $\alpha^{th}$ line segment.

As shown below, $\sigma\xi(\tau)$ mimics the process $\Gamma(\tau)$ in that both have a constant noise spectral density below a cutoff frequency. In theory, $\Gamma(\tau)$ has a constant spectral density at all frequencies, but the (fixed time-step) differential equation solver discretizes time so that $\Gamma(\tau)$ is approximated with a constant spectral density up to a cutoff related to $(1/d\tau)$. Similarly, $\xi(\tau)$ has a constant spectral density up to a cutoff related to $(1/\eta)$. So long as the cut-off is chosen above any physically relevant frequencies, $\sigma\xi(\tau)$ will provide a good estimate for $\Gamma(\tau)$.

### 3 Spectral Density

We compare the functions $\Gamma(\tau)$ and $\xi(\tau)$ through their spectral densities. $\Gamma(\tau)$ has a spectral density which is constant with frequency and has a magnitude of one [13]. We now find the spectral density of the function $\xi(\tau)$. Consider a length of time $t_f$ composed of $n$ segments of length $\eta$, such that $n = t_f/\eta$. The Fourier transform of $\xi(\tau)$ is:

$$F(\upsilon) = \int_0^{t_f} \xi(\tau) e^{i2\pi\upsilon t} d\tau \tag{5a}$$

Numerical simulation of thermal noise in Josephson circuitsNumerical simulation of thermal noise in Josephson circuits

$$= \sum_{\alpha=0}^{n-1} e^{i2\pi\upsilon\alpha\eta} \int_0^{\eta} [m_\alpha \tau + c_\alpha] e^{i2\pi\upsilon\tau} d\tau \tag{5b}$$

$$= \sum_{\alpha=0}^{n-1} e^{i2\pi\upsilon\alpha\eta} [Am_\alpha + Bc_\alpha] \tag{5c}$$

For brevity we define the complex numbers $A$ and $B$ as follows:

$$A = \frac{e^{i2\pi\upsilon\eta}(i2\pi\upsilon - 1) + 1}{-(2\pi\upsilon)^2} \tag{6}$$

$$B = \frac{e^{i2\pi\upsilon\eta} - 1}{i2\pi\upsilon} \tag{7}$$

Notice that all random terms appear in the factors $m_\alpha$ and $c_\alpha$. The spectral density is given by the magnitude squared of the Fourier Transform:

$$S(\upsilon) = \langle F(\upsilon) F^*(\upsilon) \rangle \tag{8}$$

where the brackets denote average over random outcomes and * denotes complex conjugation. Using (5) we can write this as:

$$S(\upsilon) = \sum_{\alpha=0}^{n-1} \sum_{\beta=0}^{n-1} e^{i2\pi\upsilon\eta(\alpha-\beta)} \left[ \langle m_\alpha m_\beta \rangle AA^* + \langle m_\alpha c_\beta \rangle AB^* + \langle m_\beta c_\alpha \rangle A^*B + \langle c_\alpha c_\beta \rangle BB^* \right] \tag{9}$$

The slopes and intercepts $m_\alpha$ and $c_\alpha$ of neighboring segments are correlated with each other, while those that are not neighbors are uncorrelated. To derive the correlation relations we use the property of the normal distribution and the Kronecker delta notation ($\delta_{ij} = 1$ if $i = j$, 0 otherwise):

$$\langle N_\alpha(0,1) N_\beta(0,1) \rangle = \delta_{\alpha\beta} \tag{10}$$

This gives the following correlations for $m_\alpha$ and $c_\alpha$:

$$\langle m_i m_j \rangle = \begin{cases} 2/\eta^2, & \text{if } i = j \\ -1/\eta^2, & \text{if } i = j \pm 1 \\ 0 & \text{otherwise} \end{cases} \tag{11a}$$

$$\langle m_i c_j \rangle = \begin{cases} -1/\eta & \text{if } i = j \\ 1/\eta & \text{if } i = j - 1 \\ 0 & \text{otherwise} \end{cases} \tag{11b}$$

$$\langle c_i c_j \rangle = \delta_{ij} \tag{11c}$$

Many terms in the double summation are zero and it collapses to the single sum:

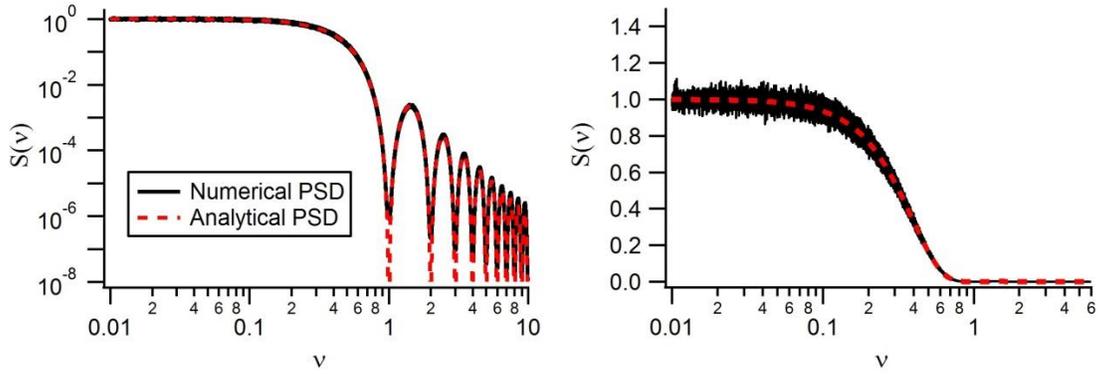

**Figure 2:** Spectral density of linearly interpolated Gaussian noise on a log-log scale (left) and a log-linear scale (right). The dotted lines represent the result of equation (13), whereas the solid lines are a numerically computed spectral density.

$$S(\nu) = \sum_{\alpha=1}^{n-2} \left\{ AA^* \frac{(2 - 2\cos(2\pi\nu\eta))}{\eta^2} + \frac{1}{\eta}\left[AB^*\left(e^{-i2\pi\nu\eta} - 1\right) + A^*B\left(e^{i2\pi\nu\eta} - 1\right)\right] + BB^* \right\} + o(n) \tag{12}$$

Here the $o(n)$ term includes the first and last terms of the sum, which are negligible for large $n$. Note for simplification that:

$$AA^* = \frac{1}{(2\pi\nu)^4}\left[2 - 2\cos(2\pi\nu\eta) + (2\pi\nu\eta)\exp\left[2\left(1 - 2\frac{\sin 2\pi\nu\eta}{2\pi\nu\eta}\right)\right]\right]$$

$$BB^* = \frac{2 - 2\cos(2\pi\nu\eta)}{(2\pi\nu\eta)^2}$$

$$2\mathrm{Re}\left[AB^*\left(e^{-i2\pi\nu\eta} - 1\right)\right] = \frac{4 - 4\cos(2\pi\nu\eta)}{(2\pi\nu\eta)^2}\left[\frac{\sin(2\pi\nu\eta)}{2\pi\nu\eta} - 1\right]$$

After simplification we obtain:

$$S(\nu) = \tau_f \eta \frac{\sin^4(\pi\nu\eta)}{(\pi\nu\eta)^4} \tag{13}$$

Figure 2 compares this analytically derived expected spectral density to a numerically computed average spectral density from 1000 linearly interpolated Gaussian noise traces. For each noise trace we use a time step of $\eta = 1$ and a number of samples $n = 65536$, such that $\tau_f = n\eta = 65536$. To compute the numerical spectral density of each trace, we sample the noise function at a sampling time of $h = \eta/32$ and use a numerical Fast Fourier Transform (FFT) procedure. Finally, we average all 1000 spectra and compare to the result of equation (13). Figure 2a shows the comparison on a log scale and one can see the agreement to several orders of magnitude; the agreement is limited by our choice of $h$ and the number of averages.

One can see that the linearly interpolated Gaussian noise is constant (white) at low frequencies. It drops to the 3-dB point (0.707) at a frequency of about $(0.23/\eta)$, close to $1/(2\pi\eta)$. As we will discuss later, $\eta$ should be chosen such that this frequency is larger than a characteristic frequency of the circuit. In making the plots for figure 2, we normalized both signals to magnitude 1 at frequency 0 by dividing by ($\tau_f \eta$). In general, the variance $\sigma^2$ of the noise should be chosen to be $1/\eta$ to mimic Gaussian white noise so that $(1/\eta)^{1/2} \xi(\tau)$ replaces $\Gamma(\tau)$. This is similar to the term, $(1/d\tau)^{1/2} N(0,1)$, used for fixed time-step solvers. The value of $1/\eta$ represents the bandwidth of the noise.

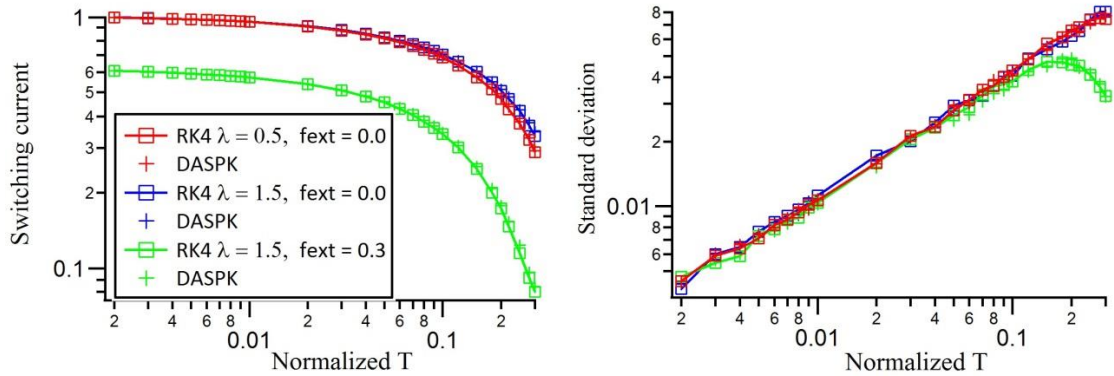

**Figure 3:** Switching current average (left) and standard deviation (right) versus normalized temperature for two different values of the inductance parameter and external frustration. Comparison is shown between the standard 4[th] order Runge-Kutta Method (RK4) and the adaptive solver DASPK.

## 4 Results on example system

We demonstrate this method by computing the switching dynamics of an underdamped DC SQUID, which consists of two identical junctions in parallel. The equations of motion are determined by current conservation and fluxoid quantization and are given by:

$$\ddot{\phi}_1 = -\gamma \dot{\phi}_1 - \sin\phi_1 + i_b/2 + \lambda(\phi_2 - \varphi_1 - 2\pi f_{ext}) + \left(2\gamma \tilde{T}/\eta\right)^{1/2} \xi_1(\tau) \quad (14a)$$

$$\ddot{\phi}_2 = -\gamma \dot{\phi}_2 - \sin\phi_2 + i_b/2 - \lambda(\phi_2 - \varphi_1 - 2\pi f_{ext}) + \left(2\gamma \tilde{T}/\eta\right)^{1/2} \xi_2(\tau) \quad (14b)$$

Here $i_b$ represents a bias current fed to both junctions, normalized to the critical current of one of the junctions. The externally applied frustration is $f_{ext} = \Phi_{ext}/\Phi_0$, with $\Phi_{ext}$ equal to the externally applied magnetic flux. The inductance parameter is $\lambda = \Phi_0/(2\pi L I_C)$, with $L$ equal to the geometric self-inductance of the loop. The subscripts 1 and 2 on $\xi(\tau)$ represent different samplings of the $\xi$ function, as the two noise sources are independent of each other.

To simulate the switching dynamics, we step the bias current in small increments of 0.01, holding for $2\times 10^3$ time units at each step. This gives a normalized ramp parameter of $di_b/d\tau = 5 \times 10^{-6}$. The normalized voltage is computed as the time average of $\dot{\phi}$. The current is ramped until the junctions switch from the superconducting state (where $\langle \dot{\phi} \rangle \approx 0$) to the finite voltage state (where $\langle \dot{\phi} \rangle \approx 1$). The current at which that jump occurs is the switching current. Thermal noise causes the jump to occur at slightly different currents each ramp. The average and standard deviation of 1000 samples are calculated at each set of parameters ($\lambda, f_{ext}, \tilde{T}$).

In the first case, we integrate equation (1) using a solver called DASPK [14], which uses a variable time-step technique, and linearly interpolated Gaussian noise for the noise term with a value of $\eta = 0.005$. In the second case, we use a fixed time-step fourth-order Runga-Kutta method with a short time-step ($d\tau = 10^{-2}$) and use a standard normal to approximate $\Gamma(\tau)$ as in equation (2). Figure 3 show the results of the comparison. The results show agreement between the two approximate solutions. The variable step solver is able to adapt the timestep without complications of resampling the noise and still attain the desired solution.

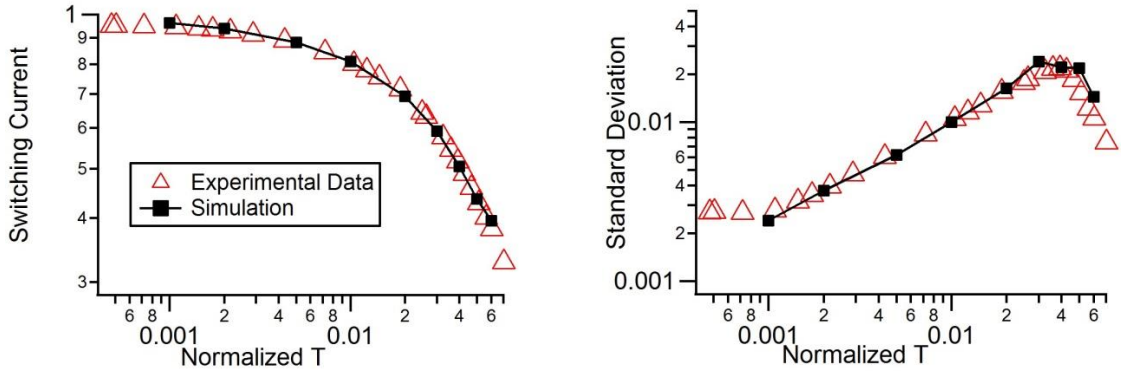

**Figure 4:** Switching current average (left) and standard deviation (right) versus normalized temperature for the data in Mannik et al. and the simulation of equations 14 a-b with $\eta = 0.025$.

## 5 Comparison to Experimental Data

We further demonstrate our method by comparing to experimental data. We fit to the data of Mannik et al. [15], where the effect of phase diffusion on the switching current is studied in moderately damped junctions. As a function of increasing temperature, the standard deviation first increases due to thermal activation, but then decreases as the system is "retrapped" and undergoes phase diffusion. This results in a peak in the standard deviation as a function of temperature [15].

The circuit model for a Josephson junction is shown in Figure 1 [15] of Mannik et al. It consists of the usual supercurrent, capacitive current and resistive current in parallel, but adds an RC-shunt to represent the high-frequency dissipative environment of the junction. This shunt gives another degree of freedom to the system, the voltage across the shunt resistor ($v_s$). Transcribing their circuit into equations of motion we obtain the following:

$$\ddot{\phi} = -\gamma\dot{\phi} - \sin\phi + i_b + v_s + \left(2\gamma\tilde{T}/\eta\right)^{1/2}\xi(\tau) \tag{15a}$$

$$\dot{v_s} = \Omega\dot{\phi}\left[v_s - \dot{\phi} - \left(2\gamma\tilde{T}/\eta\right)^{1/2}\xi(\tau)\right] \tag{15b}$$

Here $\Omega = (C/C_S)$, where $C_s$ is the shunt capacitance and $C$ is the junction capacitance as before. The normalized parameters are $\Omega = 0.017$ and $\gamma = 0.28$, computed from the junction parameters listed [15] in Mannik et al. The ramp is matched to the experiment at $di_b/d\tau = 2 \times 10^{-8}$. Figure 4 shows the comparison of the simulation and the data for the switching current and the standard deviation. Here we choose $\eta = 0.025$. The agreement is again excellent, with the simulation as well as the experiment showing the peak in the standard deviation.

## 6 Discussion

We have shown a method of numerically introducing noise into the equations of motion for a Josephson junction circuit. Our new linearly interpolated Gaussian noise, replaces the usual white noise process and can be used with variable time-step solvers. As shown in figure 3, it gives similar results to the standard numerical approach when integrating the equations of motion for a SQUID, and as shown in figure 4 it can be used to compare to experiment. Because the noise amplitude is independent of the size of the time-step of the solver, the noise function can be created ahead of time in a separate subroutine, allowing for ease of implementation. Since it can be used with variable time-step solvers, computational time can be reduced without missing important dynamics, especially for stiff systems.

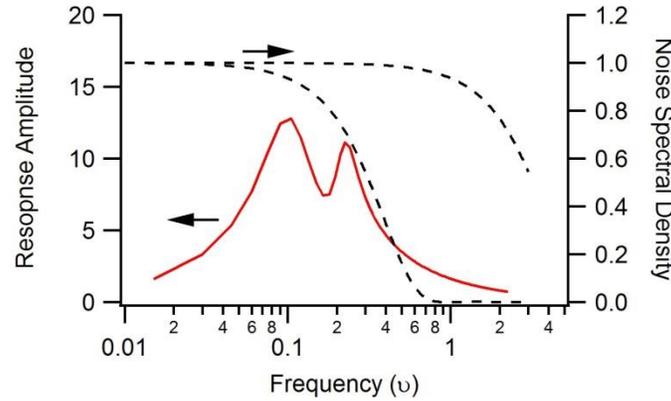

**Figure 5:** Response amplitude of the SQUID voltage (left axis, solid curve) and spectral density of the noise (right axis, dashed curves) versus the small-signal driving frequency. Two different noise spectra are shown with $\eta = 1$ and $\eta = 0.1$. The value of $\eta$ should be chosen in such a way as to keep the noise "white" at frequencies where the circuit has an appreciable response.

In using the proposed method, an important aspect is the choice of $\eta$. The value of $\eta$ should be chosen such that the cut-off frequency, $0.23/\eta$, is much larger than the frequencies at which the circuit absorbs noise. Since Josephson junctions act as nonlinear oscillators, they absorb noise most strongly at their resonant frequencies. For the system described by equations (14a-b), there are two resonant frequencies: an in-phase resonance and an out-of-phase resonance. The in-phase resonance, in the absence of damping, is given by $2\pi\upsilon = 1$. The out-of-phase resonance depends on the current, but is bounded by $2\pi\upsilon = \sqrt{2\lambda + 1}$. In figure 4, we show the voltage of the circuit when it is driven by a very small (amplitude = $0.01I_c$) AC current versus the frequency $\upsilon$ of that current. The small current is analogous to a noise current. Here we have chosen $\lambda = 0.6$ and a bias current of $1.2I_c$. Numerically, the in-phase resonance is found at $\upsilon = 0.15$ and the out-of-phase resonance is at $\upsilon = 0.225$; these match up to the predictions (0.16 and 0.23) minus a small amount due to damping. On the plot we also show the noise spectral density for $\eta = 1$ and $\eta = 0.1$. As can be seen, a choice of 0.1 is much better, since the noise is flat (white) at both resonant frequencies. This analysis is for the junctions in the zero-voltage state.

Things get slightly more complicated in the whirling (finite-voltage) state, where the junction phases change continuously with time. In such a case the system becomes described by a damped and driven Mathieu equation, where the noise acts as the driving term [16]. The undriven Mathieu equation can display parametric instabilities. For a SQUID or a parallel array, these frequencies are again bounded by an upper frequency given by $2\pi\upsilon = \sqrt{2\lambda + 1}$ [17]. The locations of the exact resonances are hard to predict, but for the purposes of choosing $\eta$, only the largest frequency matters.

In general, there is no definitive rule for the choice of $\eta$; smaller $\eta$ will give more accurate results, but with diminished effect as $\eta$ goes to zero. The advantage of our method is that once

Numerical simulation of thermal noise in Josephson circuitsthe highest resonant frequency of the circuit has been determined, either analytically or numerically like in figure 4, the choice of $\eta$ can be fixed. If the circuit is driven at different frequencies, for example through the application of NMR-style pulses for quantum coherence experiments, the value of $\eta$ does not need to be changed. The variable time-step solver will adjust to the faster drive and dynamics will not be missed.

## Acknowledgements

We thank J.J. Mazo for useful discussions. Funding for one of us (KS) was partially provided by NSF DMR 0509450.## References

[1] Fulton T and Dunkleberger L 1974 *Physical Review* **B9** 4760
[2] Devoret M, Martinis J and Clarke J 1985 *Physical Review Letters* **55** 1908
[3] Kautz R and Martinis J 1990 *Physical Review* **B42** 9903
[4] Devoret M H and Schoelkopf R J 2013 *Science* **339** 1169
[5] Orlando T P, Mooij J E and van der Zant H J 1991 *Physical Review* **B43** 10218; Resnick D J, Garland J C, Boyd J T, Shoemaker S S and Newrock R S 1981 *Physical Review Letters* **47** 1542
[6] Binder P, Abraimov D, Ustinov A, Flach S and Zolotaryuk Y 2000 *Physical Review Letters* **84** 745; Trias E, Mazo J and Orlando T 2000 *Physical Review Letters* **84** 741
[7] Clarke J and Braginski A eds. 2004 *The SQUID Handbook Vol. I* (Wiley-VCH)
[8] Siddiqi I, Vijay R, Pierre F, Wilson C, Frunzio L, Metcalfe M, Rigetti C, Schoelkopf R, Devoret M, Vion D and Esteve D 2005 *Physical Review Letters* **94** 027005
[9] Lisenfeld J, Lukashenko A, Ansmann M, Martinis J, and Ustinov A 2007 *Physical Review Letters* **99** 170504
[10] Orlando T and Delin K 1991 *Foundations of Applied Superconductivity* (Prentice Hall)
[11] Ascher U M and Petzold L R 1998 *Computer Methods or Ordinary Differential Equations and Differential-Algebraic Equations* (SIAM)
[12] Cash J R and Karp A H 1990 *ACM Transactions on Mathematical Software* **16** 201
[13] Gillespie D T 1996 *American Journal of Physics* **64** 225
[14] Van Keken P E, Yuen D A and Petzold L R 1995 *Geophysical and Astrophysical Fluid Dynamics* **80** 57
[15] Mannik J, Li S, Qiu W, Chen W, Patel V, Han S and Lukens J E 2005 *Physical Review* **B71** 220509
[16] Nayfeh A H and Mook D T 1979 *Nonlinear Oscillations* (Wiley and Sons)
[17] Watanbe S, Strogatz S, van der Zant H, and Orlando T P 1995 *Physical Review Letters* **74** 379